\documentclass[10pt]{article}
\usepackage{a4}
\usepackage{graphicx}

\begin{document}

\title{The nuclear density of states and the role of the residual interaction}
\author{Calvin W. Johnson and Edgar Ter\'an, \\ Department of Physics, San Diego State University \\
5500 Campanile Drive, San Diego, CA 92182-1233 \\ USA }
\date{  }
\maketitle

\begin{abstract}
We discuss the role of mean-field and moment methods in
microscopic models for calculating the nuclear density of states
(also known as the nuclear level density). Working in a shell-model framework,
we use moments of the nuclear many-body Hamiltonian to
illustrate the importance  of the residual interaction for accurate
representations.
\end{abstract}

\section{Introduction: mean-fields and moments}

``\textit{We all know that Art is not truth. Art is a lie that makes
us realize the truth, at least the truth that is given to us to
understand.}'' --Pablo Picasso.

If you want to model nuclear reactions in a hot, dense, neutron-rich
environment, whether for astrophysics or for stockpile stewardship,
you need to compute neutron capture rates where you  have a thick forest
of states or resonances\cite{RTK97}. If you want to compute statistical neutron capture
using the Hauser-Feshbach formalism\cite{HF52}, you need the density of states, or
the nuclear level density \cite{NLD}.  The nuclear level density is a challenge
both experimentally and theoretically, because in essence one needs a
reliable count of thousands or millions of states. The level density is
generally quoted as being the most uncertain input into statistical capture
calculations\cite{RTK97}.

You don't need information on each individual state or resonance; rather you need
to average over many many states. The tools to average over the states, or rather over the
nuclear many-body Hamiltonian, are mean-field methods and moment methods; later on
we will discuss how moment methods can be considered as a generalization of
mean-field methods.

A static mean-field picture is the basis for the most widely used approach to
level densities, generalized Bethe Fermi gas models\cite{Be36,DSVU73}. Here one starts with a
single-particle spectrum generated by a mean-field and derives the density of
many-body states. The most sophisticated applications extract the single-particle
spectrum from mean-field calculations, such as Skyrme Hartree-Fock\cite{Go96}.  The residual
interaction is mostly ignored, although pairing interactions are included in part
through a quasi-particle spectrum from Hartree-Fock-Bogoliubov, and corrections can be
added for rotational motion, etc.. Although such approaches are simplest to use,
the lack of consistent use of the entire residual interaction is arguably unsatisfactory.

The best framework to fully include the residual interaction is that of the
interacting shell model, which uses a basis of many-body Slater determinants.
For application to low-energy spectroscopy, one
typically diagonalizes the nuclear Hamiltonian, but this is impractical for
useful applications to the level density.

\section{Spectral distribution methods}

Nuclear statistical spectroscopy starts from the moments of the
Hamiltonian. For example, in a finite space, most
state densities of many-body systems with a two-body interaction
tend toward a Gaussian shape \cite{MoFr75}
characterized by the first and
second moments of the Hamiltonian. In most realistic cases the
density has small but non-trivial deviations from a Gaussian, so one
requires higher moments.

None of the formalism in this section
is original; a thorough reference to statistical
spectroscopy is Ref.~\cite{Wo86}, although some of our notation is different.
Due to space limitations we give only a brief overview; for more
careful discussion, readers are referred to our preprint\cite{TeJo05}.

 We work in a finite model space
${\cal M}$ wherein the number of particles is fixed. If in ${\cal
M}$ we represent the Hamiltonian as a matrix $\mathbf{H}$, then all
the moments can be written in terms of traces.
The total dimension of the space is $D = \mathrm{tr\,}\mathbf{1}$, and the
average is
$\langle \mathbf{O} \rangle = D^{-1} \mathrm{tr \,} \mathbf{O}.$
The first moment, or centroid, of the Hamiltonian is
$\bar{E} = \langle \mathbf{H} \rangle;$
all other moments are \textit{central} moments, computed relative to
the centroid:
\begin{equation}
\label{eq:shell_moms}
\mu^{(n)} = \langle (\mathbf{H}-\bar{E})^n \rangle \; \; .
\end{equation}

The width $\sigma$ is given by $\sqrt{\mu^{(2)}}$, and one scales
the higher moments by the width:
\begin{equation}
m^{(n)} = \frac{\mu^{(n)}}{\sigma^n}.
\end{equation}

In addition to the centroid and the width, the next two moments have
special names. The scaled third moment $m^{(3)}$ is the
\textit{asymmetry}, or the skewness.

It has long been realized that, rather than
computing higher and higher moments of the total Hamiltonian, one
could partition the model space into suitable subspaces and compute
just a few moments in each subspace. In particular, if the space is 
partitioned using spherical shell-model
configurations, that is, all states of the form
$(0d_{5/2})^4(1s_{1/2})^2 (0d_{3/2})^2$, etc., then it is possible
to derive expressions for the configuration moments directly in
terms of the single-particle energies and two-body matrix elements,
without constructing any many-body matrix elements.

We use $\alpha,
\beta, \gamma, \ldots$ to label subspaces. Let
\begin{equation}
P_\alpha =\sum_{i \in \alpha} \left | i \right \rangle \left \langle
i \right |
\end{equation}
be the projection operator for the $\alpha$-th subspace.
One can introduce \textit{partial} or \textit{configuration densities},
\begin{equation}
\rho_\alpha(E) = \mathrm{tr} \, P_\alpha \delta(E-\mathbf{H} ).
\end{equation}
The  total level density $\rho(E)$ is just the sum of the
partial densities.

Now we
define \textit{configuration moments}: the configuration dimension is
$D_\alpha = {\rm tr \,} P_\alpha$, the configuration centroid is
$\bar{E}_\alpha = D_\alpha^{-1} {\rm tr \,} P_\alpha \mathbf{H},$
while the configuration width $\sigma_\alpha$ and configuration
asymmetry $m_\alpha^{(3)}$ are defined in the obvious ways.

One important result from spectral distribution theory is that the
configuration centroids depend entirely upon the single-particle
energies and the monopole-monopole part of the residual
interaction\cite{DuZu99}.
The monopole interaction is attributed to mean-field and saturation
properties of the nuclear interaction. One can subtract out the
monopole interaction exactly, which sets the centroids to zero but
leaves the widths unchanged; such a monopole-subtracted interaction
is referred to as a ``traceless'' interaction.
 For some more information on computing the
monopole interaction, see the references \cite{Wo86,TeJo05}.

\section{First moments: the mean field}

\begin{figure}
\begin{center}
\includegraphics[scale=.5]{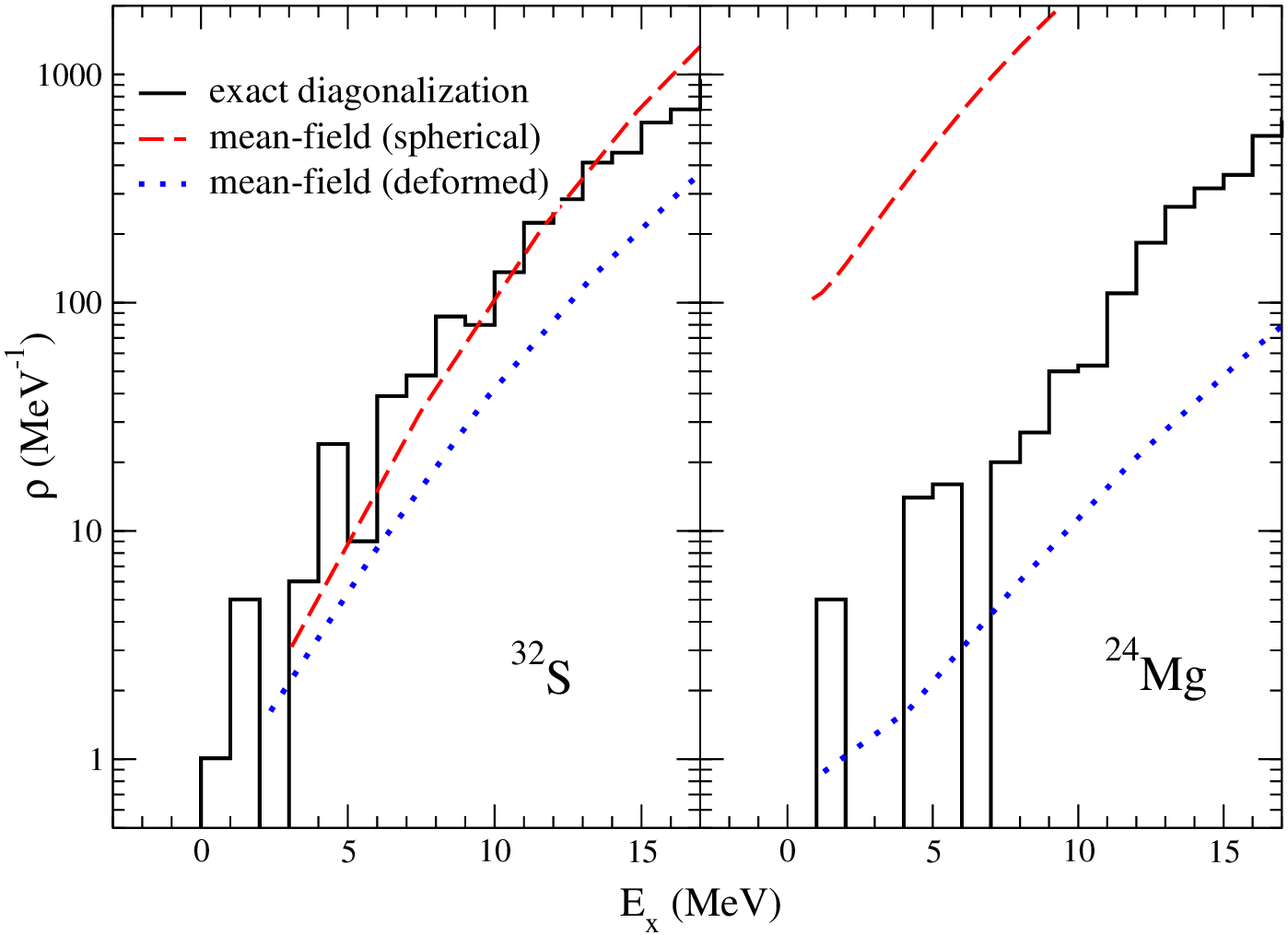}
\end{center}
\caption{Comparison of exact level density (from complete
diagonalization of shell-model Hamiltonian) with Fermi gas model,
using spherical (dashed lines) mean-field or deformed (dotted)
mean-field. Left: $^{32}$S. Right: $^{24}$Mg. Both are computed in
$sd$-shell with USD interaction. \label{meanfield}}
\end{figure}

The \textit{thermodynamic} approach to the density of states is to
note that the partition function $Z$ is the Laplace transform of the
density, $ Z(\beta) = \int e^{-\beta E} \rho(E) dE $. The inverse
Laplace transform is accomplish through either a saddle-point
approximation \cite{Be36,NA97} 
or maximum entropy \cite{Or97}. The many-body partition function
can in principle be computed  from the full Hamiltonian, including
the residual interaction, using path integrals\cite{NA97,Or97},
but the integrals must be evaluated through Monte Carlo sampling and
in order to avoid the sign problem, only a restricted class of
Hamiltonians can be used\cite{LJKO93}.

Instead, as mentioned in the introduction, the standard approach
is to approximate the many-body partition
function using non-interacting particles in a mean field\cite{Be36}. That is,
\begin{equation}
\ln Z(\alpha,\beta) = \int g(\epsilon)\ln \left( 1+ \exp(\alpha -
\beta \epsilon )\right) d\epsilon
\end{equation}
where $g(\epsilon) = \sum_i \delta(\epsilon - \epsilon_i )$ is the
single-particle density of states, $\{ \epsilon_i\}$ are the
mean-field single-particle energies, and $\alpha$ is proportional to
the chemical potential. The excitation energy and the particle
number are set through the derivatives of $\ln Z$ with respect to $\beta$ and 
$\alpha$, respectively. 

For an example of applying this method, Goriely \textit{et al}\cite{Go96} 
extract single-particle energies from
Skyrme Hartree-Fock calculations and use it to estimate level
densities throughout the nuclear landscape. To illustrate this, we
consider a nontrivial model system, a finite-basis interacting shell
model, where we can compare the ``true'' density of states (from
numerical diagonalization of the many-body Hamiltonian) with
approximate methods. We work in the $sd$-shell and lower $pf$-shell
where there are sufficiently few number of levels (a few thousand)
so that one can carry out the full diagonalization using 
the shell model codes OXBASH\cite{OXBASH} and REDSTICK\cite{REDSTICK}. The Hartree-Fock
calculations were done using a shell-model based code, SHERPA, used
to test the random phase approximation in exactly the same way\cite{SJ02}.

Two sample results are shown in Fig.~(\ref{meanfield}), for $^{32}$S
and $^{24}$S, with valence nucleons in the $sd$-shell using 
the so-called universal \textit{sd}-shell (USD) interaction\cite{Wi84}. 
The Fermi gas calculations are
done exactly according to the prescription with no corrections.

One sees that for a spherical nucleus, $^{32}$S, the mean-field
Fermi gas calculation works extremely well for such a simple
prescription. $^{24}$Mg, which is deformed, is much more
problematic. For both nuclides we computed the single-particle
energies for both spherical and deformed mean-fields. Neither yield
a very good description for $^{24}$Mg (and for other deformed
nuclei), and indeed in practical applications\cite{Go96} one
typically includes ``rotational enhancement'' and other corrections.

The sample calculation in Fig.~(\ref{meanfield}) is crude and should
not be taken as an indictment of Fermi gas model calculations; 
lacking other approaches, they are the best game in
town. The above exercise is intended to inculcate a desire to 
go beyond the mean-field to include the residual interaction.

\section{Second moments: the residual interaction}

To go beyond the mean field and include the residual interaction, we
turn to the concepts of Section 2.  In particular we write the total
density as a sum of configuration densities. But how do we model the
configuration densities?

\begin{figure}
\begin{center}
\includegraphics[scale=.5]{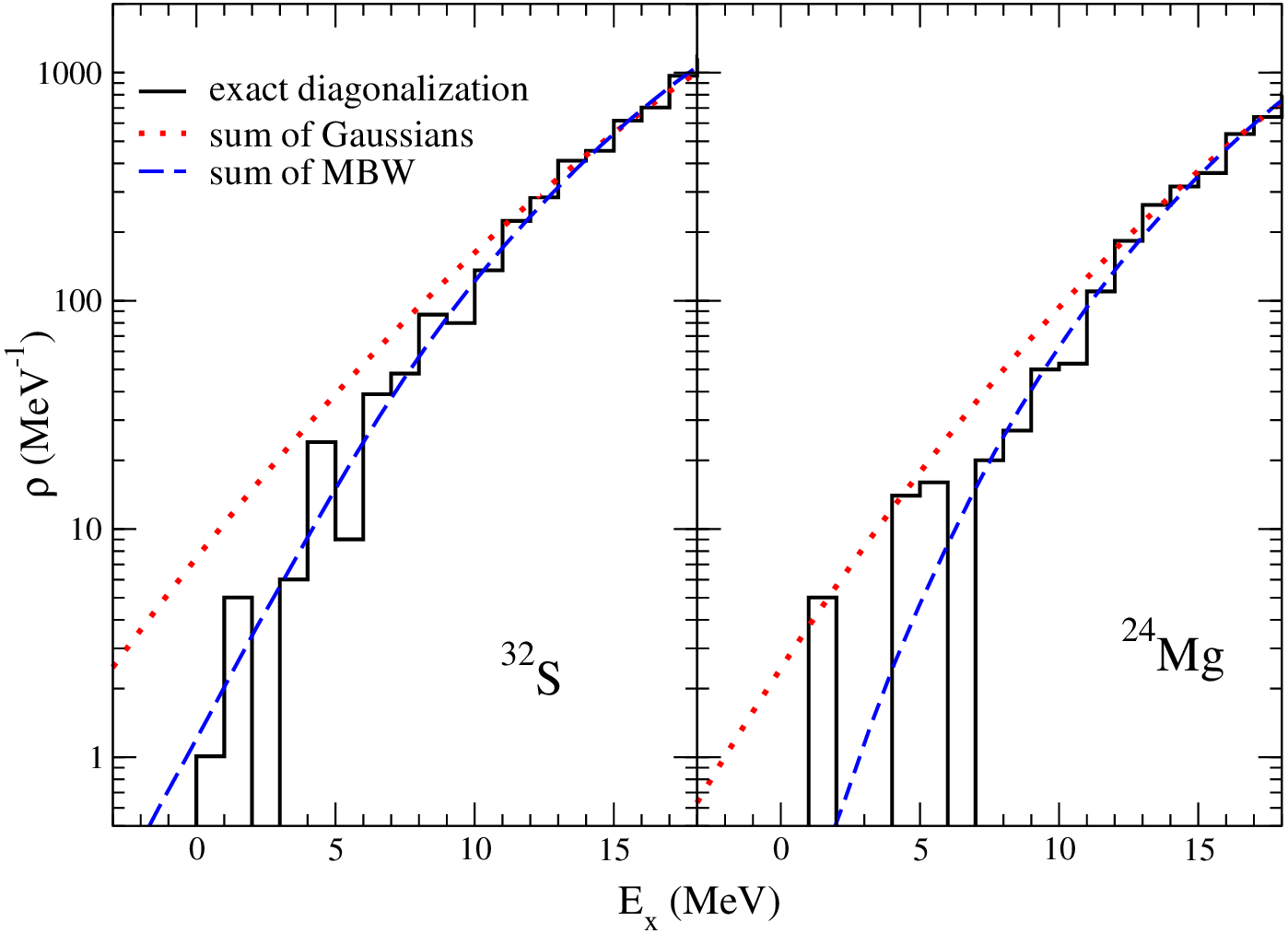}
\end{center}
\label{gauss_mbw} \caption{Comparison of exact level density (from
complete diagonalization of shell-model Hamiltonian) with sum of
Gaussians.}
\end{figure}

In this section we approximate the configuration densities as
Gaussians\cite{HKZ03}, with the average and width given by the configuration
centroid and width, respectively. We remind the reader that the
configuration centroids arise from the mean-field, while the width, or 
second moment, 
reflects the strength of the residual interaction. Elsewhere 
we show that the configuration width is approximately constant\cite{TeJo05}, at
least within a major oscillator shell.

The results  for $^{32}$S and $^{24}$Mg are plotted in 
Fig.~(2) (dotted line). While for the latter this is
an improvement over the zeroth order mean-field results as seen in
Fig 1, it still does poorly at low energy. (For some nuclides, not shown 
here, the Gaussian approximation is sufficient\cite{HKZ03}.) Therefore we must go
beyond second moments and go to \textit{third} moments.

\section{Third moments: collectivity}

First configuration moments (centroids) 
reflect the mean-field; second moments (widths)
reflect the overall strength of the residual interaction. Third moments, 
we claim, are indicative of collectivity. The argument follows from 
simple pictures of collectivity, whereby one has a single collective state 
pushed down or up in energy relative to the remaining degenerate noncollective states.
Because of the inherent asymmetry one has a nontrivial third moment. 
This is distinct from the spreading width (second moment) which, in simplest 
terms, spreads out the states symmetrically. 
Calculations of well-known collective interactions, such as $Q\cdot Q$, 
show strong persistent third moments \cite{TeJo05}. The real situation is 
 more subtle than this, but as a simple paradigm of 
level densities this argument suffices. 

To include third moments in our models for partial densities, 
we need something beyond Gaussians for the
configuration densities. Many suggestions have been made, such as
Gram-Charlier expansions\cite{Wo86},
Cornish-Fisher expansions\cite{KPS86}, or binomials\cite{Zu01}. We use
a modified Breit-Wigner (MBW) form,
\begin{equation}
\rho_\mathrm{MBW}(E) = \frac{1}{W^3}\frac{ (E_\mathrm{max} -
E)^2(E-E_\mathrm{min})^2}{(E-E_0)^2 +W^2} \label{mbw}
\end{equation}
which has several advantageous characteristics, including
positive-definite on the interval $E_\mathrm{min} < E <
E_\mathrm{max}$, and exact analytic moments.  The four parameters of
Eq.~(\ref{mbw}) can be fitted to reproduce the first four
configuration moments.

Fig.~(2) compares exact calculations of $^{32}$S
and $^{24}$Mg with a sum of MBW configuration densities (dashed
line), which are significant improvements over the sum of Gaussians
(dotted lines). For this calculation we computed the third and fourth configuration
moments from the many-body Hamiltonian matrix, generated using the 
REDSTICK shell model code. In principle, one can compute the moments directly from
the two-body matrix elements\cite{Wo86} but we have found apparently
discrepancies between the published formulas and direct calculation,
which we have been unable to resolve so far. A more serious problem
is that even if one would able to compute the third and fourth
moments directly, they are very time-consuming for large spaces. Two 
options suggest themselves. First, as seen in
Fig.~(2), at moderate excitation energy the Gaussian
and MBW curves converge, suggesting that one only needs 3rd and 4th
moments at lower energy. This further supports our suggestion to
think of the third moment as a sign of collectivity. Second, at least 
part of the asymmetry in fact arises from the mean-field, which leads to 
approximate formulas for the asymmetry\cite{TeJo05}. Unpublished results
suggest these approximations are partially but not wholly successful, so 
more work is required.

\section{Comparision with experiment}

So far we have only compared models against models, which in our case
means validating an approximate approach against a more detailed
calculation, using exactly the same input. But we need to compare
against real experimental data. While mean-field/Fermi gas models
have been compared extensively against experiment \cite{DSVU73,Go96}, moments
methods have rarely been compared against experiment\cite{PGM00}.

\begin{figure}[t]
\begin{center}
\includegraphics[scale=.5]{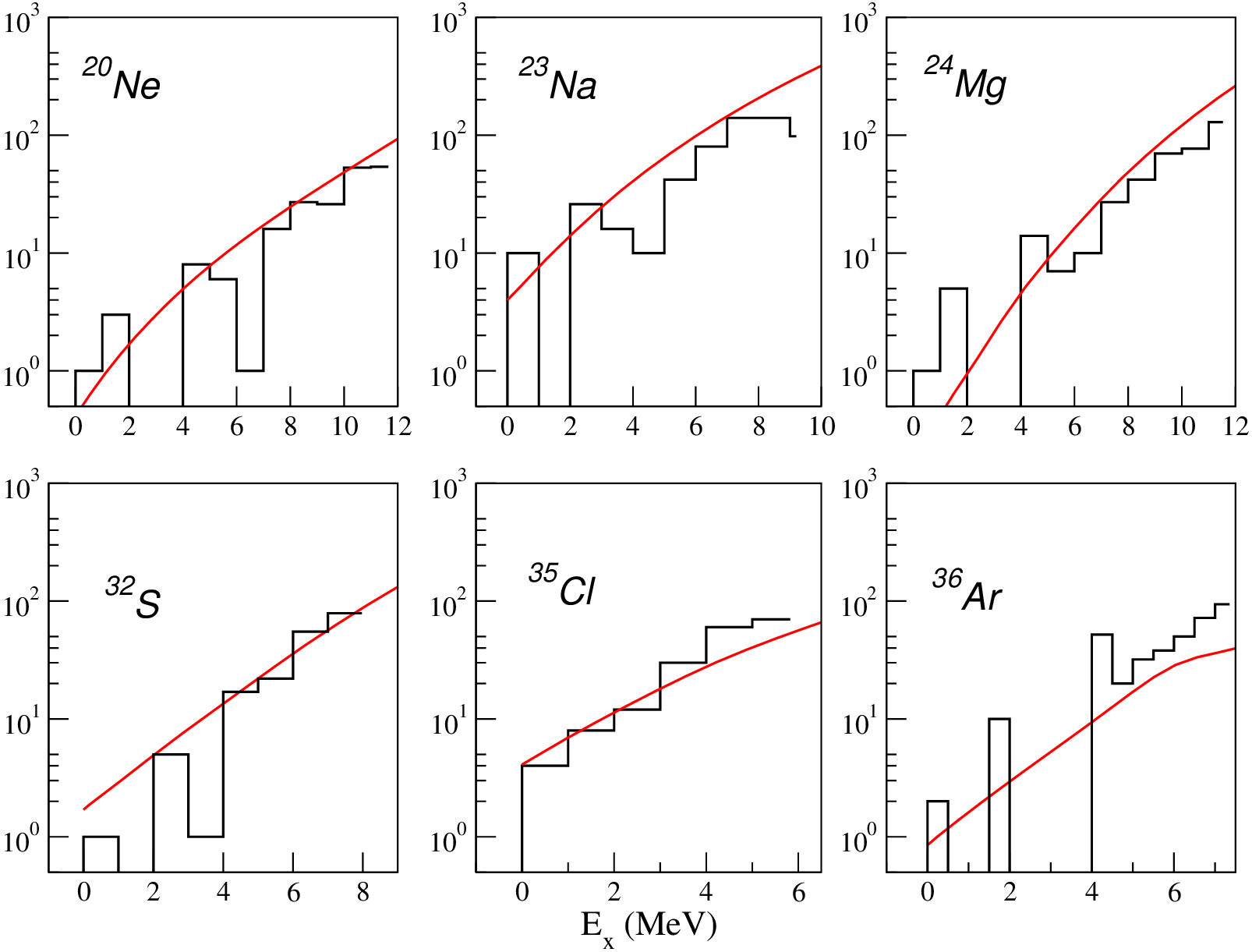}
\end{center}
\label{expt} \caption{Comparison of experimental state 
densities  (binned) of 
$sd$ shell nuclides against our calculations, 
including factor-of-two for opposite parity states.}
\end{figure}

Although important technical issues remain (such as accurate and
efficient calculation of third and fourth moments when needed, and
determination of the ground state energy), for large-scale
application of moments methods the most significant issue is the
choice of interaction. Most shell-model interactions are adjusted
for relatively small model spaces; but believeable calculations of
the density of states requires so-called ``intruder'' configurations
such as those of opposite parity. This will be the major focus of
subsequent work.

Nonetheless we can perform preliminary calculations. We show them in
Figs.~(3) ($sd$-shell using the USD interaction\cite{Wi84}) 
and (4) ($pf$-shell using GXPF1\cite{gxpf1}).  
The experimental data are taken from individual states obtained 
from the Reference Input Library of the Nuclear Data Services site \cite{NDS}. 
The third and fourth moments were computed in an
approximate fashion. The ground state energies were computed either
using REDSTICK or via Hartree-Fock + RPA (the code SHERPA)\cite{SJ02}. 
As the model spaces were for a single parity, we multiplied by a factor of 2 to estimate the
opposite parity contribution. This is a crude approach and one which
needs refinement; there have been many papers examining the relative
contribution of opposite parity states to the level density\cite{parity}.
We do well for $sd$-shell nuclides and, in the $pf$ shell, for $^{48}$Ti 
and $^{45}$Sc, but not so well for $^{51}$V and 
$^{52}$Cr. While our crude treatment of parity may be partly at fault, 
the most likely culprit is the interaction, and in particular the 
monopole (mean-field) structure of the interaction.

\begin{figure}[t]
\begin{center}
\includegraphics[scale=.5]{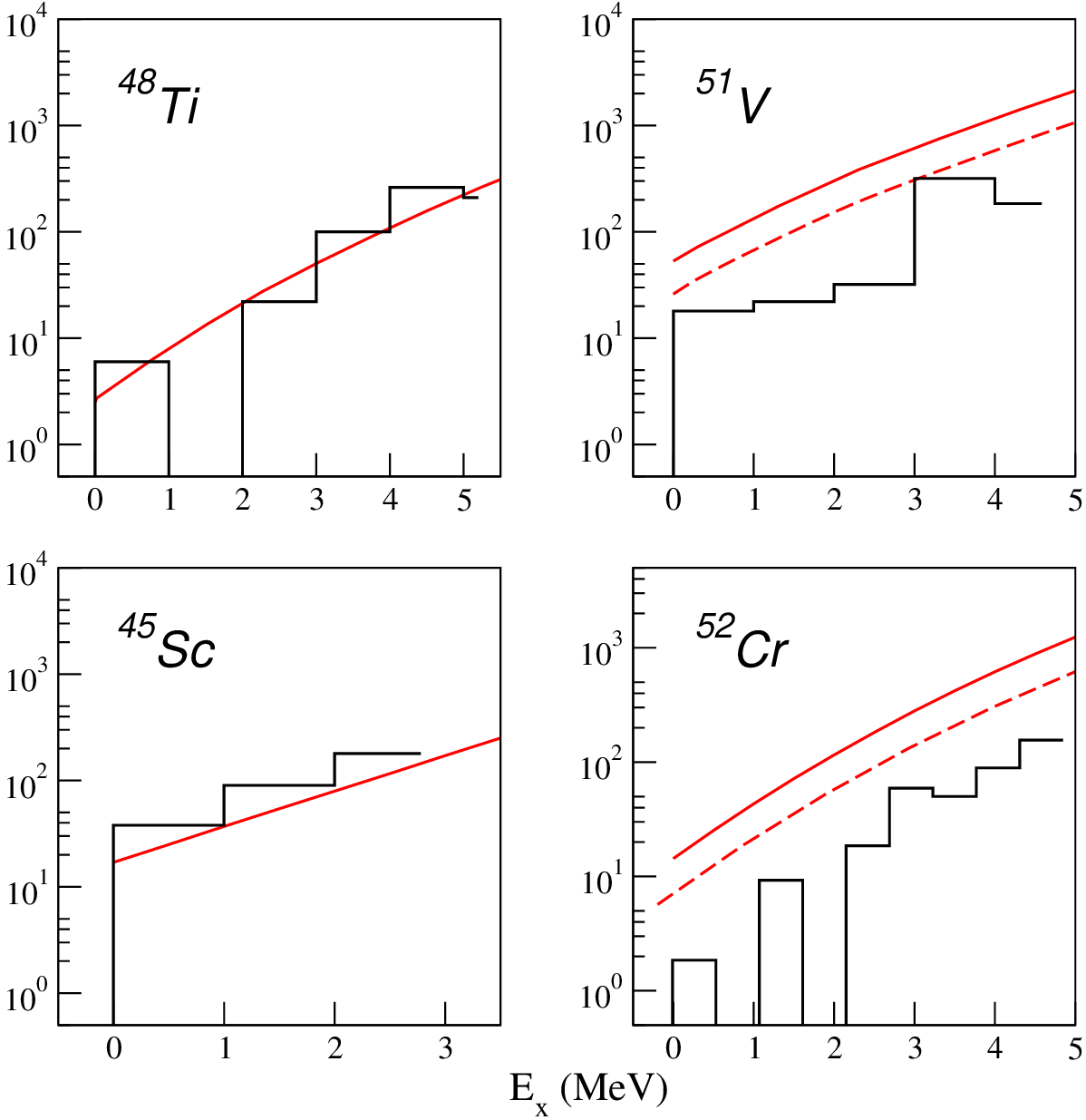}
\end{center}
\caption{Same as Fig.~3, but in $pf$ shell. Dashed lines 
on $^{52}$V and $^{52}$Cr are for normal parity states only.}
\end{figure}

\section{Conclusions}

We have reviewed applying spectral distribution methods to the
nuclear density of states. To guide the reader, we linked the first moment to
Hartree-Fock energies and effective single-particle energies, which
form the basis for mean-field Fermi gas calculations. Use of the
residual interaction lead to higher moments, in particular the
second and third moments, which we interpret as spreading widths 
and collectivity, respectively.

Finally, we can suggest a revision of Picasso's quote:

\textit{We all know that models are not reality. Models are lies
that makes us understand reality, at least that part of reality that
is given to us to understand.}

This work is supported by grant DE-FG52-03NA00082  from the
Department of Energy /National Nuclear Security Agency.  CWJ
acknowledges helpful conversations with Dr. W. E. Ormand.


\begin{thebibliography}{99}
\bibitem{NLD} The state density includes $2J+1$ degeneracies, while
the level density does not. For applications it is important to distinguish
between them, but, following colloquial usage, we use the terms interchangeably.
Except for $J$-projected level densities, everything else we discuss is
technically a state density.
\end{thebibliography}
\end{document}